\documentclass[runningheads]{llncs}
\usepackage{graphicx}
\usepackage{multirow}
\usepackage{array}
\usepackage{amssymb}
\usepackage{mathrsfs}
\usepackage{bbm} 
\usepackage{hyperref}

\begin{document}
%
% \title{Contribution Title\thanks{Supported by organization x.}}
\title{Reliable Source Approximation: Source-Free Unsupervised Domain Adaptation for Vestibular Schwannoma MRI Segmentation}

\author{
Hongye Zeng\inst{1}\and
Ke Zou\inst{2}\and
Zhihao Chen\inst{3}\and
Rui Zheng\inst{1,4} \and
Huazhu Fu\inst{5}
}
\authorrunning{Zeng et al.}
\institute{
School of Information Science and Technology, ShanghaiTech University, Shanghai, China \and
National Key Laboratory of Fundamental Science on Synthetic Vision, College of Computer Science, Sichuan University, Chengdu, China \and
College of Intelligence and Computing from Tianjin University, China \and
Shanghai Engineering Research Center of Intelligent Vision and Imaging, Shanghai, China \and
Institute of High Performance Computing (IHPC), Agency for Science, Technology and Research (A*STAR), Singapore 138632, Republic of Singapore\\
\email{\{zenghy,zhengrui\}@shanghaitech.edu.cn,\ hzfu@ieee.org}\\
}
\renewcommand{\thefootnote}{}
\footnotetext{This work is supported by the Natural Science Foundation of China (No. 12074258, hosted by Rui Zheng), and Huazhu Fu’s Agency for Science, Technology, and Research (A*STAR) Central Research Fund.\\
Rui Zheng and Huazhu Fu are the corresponding authors.}
\maketitle    
\begin{abstract}
\begin{sloppypar}
Source-Free Unsupervised Domain Adaptation (SFUDA) has recently become a focus in the medical image domain adaptation, as it only utilizes the source model and does not require annotated target data. However, current SFUDA approaches cannot tackle the complex segmentation task across different MRI sequences, such as the vestibular schwannoma segmentation. To address this problem, we proposed Reliable Source Approximation (RSA), which can generate source-like and structure-preserved images from the target domain for updating model parameters and adapting domain shifts. Specifically, RSA deploys a conditional diffusion model to generate multiple source-like images under the guidance of varying edges of one target image. An uncertainty estimation module is then introduced to predict and refine reliable pseudo labels of generated images, and the prediction consistency is developed to select the most reliable generations. Subsequently, all reliable generated images and their pseudo labels are utilized to update the model. Our RSA is validated on vestibular schwannoma segmentation across multi-modality MRI.  The experimental results demonstrate that RSA consistently improves domain adaptation performance over other state-of-the-art SFUDA methods. Code is available at \href{https://github.com/zenghy96/Reliable-Source-Approximation}{https://github.com/zenghy96/Reliable-Source-Approximation}.
% \url{https://github.com/zenghy96/Reliable-Source-Approximation}.
\end{sloppypar}

\end{abstract}
\section{Introduction}
Supervised learning models have recently made substantial advancements in medical image analysis and typically assume the same data distribution between train and test data \cite{estevaDeep2021}. This assumption challenges them in real-world practice when confronted with data collected from different imaging systems \cite{redkoAdvances2019}. To address this domain-shift problem, many adaptation approaches have been proposed and their application scenarios are becoming increasingly practical. Unsupervised domain adaptation (UDA) \cite{ganin2016domain,longConditional2018} is gradually being replaced by source-free unsupervised domain adaptation (SFUDA) \cite{liangWe2021} since source domain data is often invisible due to privacy concerns.

Entropy minimization is a popular technique in existing SFUDA approaches and aims to produce more confident model predictions \cite{BatesonSFDA,batesonSourcefree2022a}. TENT \cite{wangTent2021} is the pioneer in introducing this concept, which minimizes the entropy of model predictions to reduce generalization error. TENT necessitates continuous updates to the model, which potentially incurs significant computational costs and suffers from unstable performance due to small batch size and imbalanced data. ETTA \cite{niuEfficient2022} needs lower computation costs by identifying reliable and non-redundant samples. SAR \cite{niuStable2023} explores obstacles that harm the model performance. Pseudo-labeling is another mainstream SFUDA technique and aims to discard or correct erroneous pseudo-labels, which inevitably appear under the influence of domain shifts. Currently, there exist three distinct solutions: enhancing the quality of pseudo-labels via denoising, screening out inaccurate pseudo-labels, and devising a robust disparity metric for pseudo-labels. Many SFUDA methods directly use the class with the highest prediction probability as the pseudo-label \cite{zhou2022domain,prabhu2022augco,kothandaraman2021sssfd,dasgupta2022overcoming,tian2023robust,hou2021visualizing}. To reduce the noise pseudo-labels generated by the argmax operation, several approaches primarily focus on developing diverse filtering mechanisms to utilize dependable pseudo-labels selectively, such as maximum prediction probability \cite{zhou2022domain,prabhu2022augco,kothandaraman2021sssfd,dasgupta2022overcoming}, self-entropy \cite{tian2023robust}, consistency score \cite{prabhu2022augco,hou2021visualizing}, and weight-averaged predictions \cite{wangContinual2022}.
However, most SFUDA methods are validated on domain-shift data caused by different centers, protocols, or devices, and are unable to effectively handle the segmentation task between different MRI sequences, which have relatively large domain shifts.
% as shown in Figure \ref{fig:intro}. 

Complex medical image segmentation tasks in domain adaptation commonly use source approximation, which aims to generate source-like images and is generally implemented in UDA scenarios through CycleGAN \cite{zhuUnpaired2020}. Cai et al. \cite{Cai2019Towards} introduce a cross-modality MR/CT segmentation framework utilizing CycleGAN, which integrates a shape consistency loss to maintain anatomical coherence across source and target images. Jiang et al. \cite{Jiang2019Integrating,Jiang2018TumorAware} employ CycleGAN to produce MRI images from CT scans for lung tumor segmentation, facilitating precise delineation of tumors adjacent to soft tissues.
For SFUDA scenarios, source approximation is implemented by Fourier transform. For example, FSM \cite{yangSource2022} acquires a preliminary source image by immobilizing the source model and refining a trainable image, which is enhanced through mutual Fourier Transform. The resultant refined source-like image depicts the source data distribution, thereby aiding in domain alignment throughout the adaptation procedure.

first attains a coarse source image by freezing the source model and training a learnable image, then refines the image via mutual Fourier Transform. The refined source-like image provides a representation of the source data distribution and facilitates domain alignment during the adaptation process.

% 1. 给方法起一个名字，SFUDA，我看前面也是SFUDA方法，还有你文章里的缩写感觉太多了，到时候被审稿人说体验不好，读起来不通顺。
% 2. uncertainty refinement 啥意思啊？感觉不是特别好懂，你就说我们引入了不确定性估计来做筛选可信的预测？

In this paper, we proposed the Reliable Source Approximation (\textbf{RSA}) for vestibular schwannoma segmentation across different MRI sequences. In contrast to previous SFUDA source approximations, our method innovatively uses edge-guided image translation to generate source-like and structure-preserved images from the target domain. To obtain the most reliable generated images and their pseudo-labels, an uncertainty segmentation model is introduced to produce reliable pseudo-labels through uncertainty refinement, and the prediction consistency is proposed to identify the optimal approximation result. 
Our contributions can be summarized as follows: (1) We propose a novel source approximation method through edge-guided image translation. (2) We introduce the uncertainty refinement and prediction consistency to obtain the most reliable approximation result. (3) Our proposed RSA outperforms other SFUDA methods on adaptive vestibular schwannoma segmentation.

\section{Methodology}
\begin{figure}[t]
    \centering
    \includegraphics[width=\textwidth]{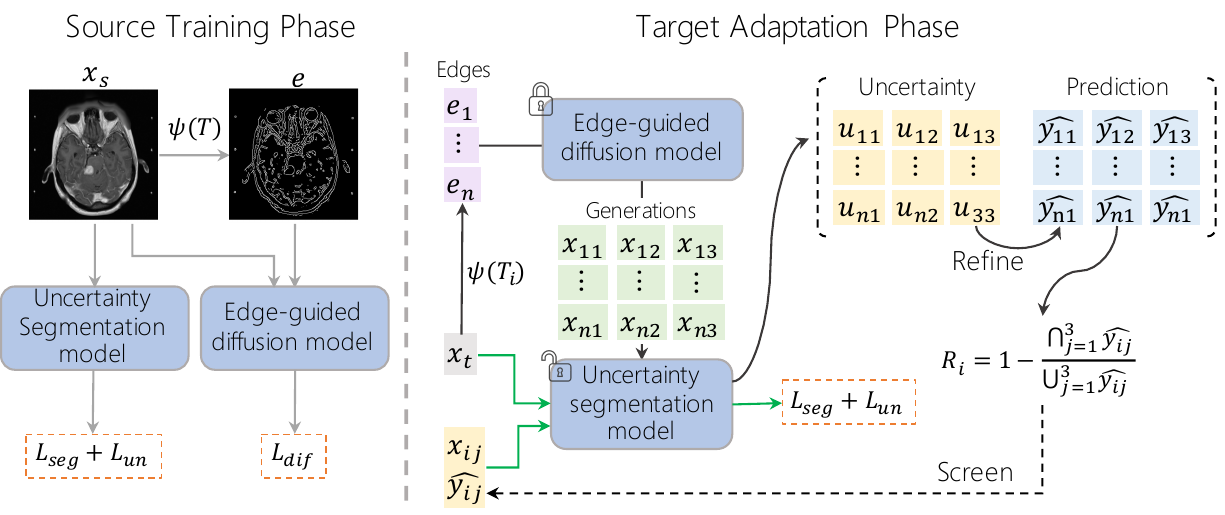}
    \caption{Overview of the proposed SFUDA method. Two models are pre-trained using source data, and the target adaptation phase consists of Reliable Source Approximation (\textbf{RSA}, black arrows) and model fine-tuning (green arrows). }
    \label{fig:method}
\end{figure}

Figure \ref{fig:method} illustrates our SFUDA framework via reliable source approximation. Given the source data with annotation $D_s=\{x_s^i,y_s^i\}_{i=1}^N$, we first train the edge-guided diffusion model $g_\theta$ and the uncertainty segmentation model $f_\theta$. Then, the source-like and structure-preserved images are generated under spatial guidance from edges. The prediction consistency and uncertainty are used to find the optimal generated image $x_{ij}$ and its reliable pseudo label $\hat{y_{ij}}$. Finally, all $x_t$, $y_{ij}$, and $\hat{y_{ij}}$ are used to finetune the uncertainty segmentation model for adaptation.

\subsection{Edge-guided Image Translation}
The edge-guided image translation aims to translate image style and simultaneously preserve the original spatial structure. To achieve this, we first train a basic denoising diffusion probabilistic model (DDPM) \cite{hoDenoising2020} for source image synthesis. It can generate source-like images by progressive denoising an input Gaussian noise, but generated images are completely random without any spatial guidance. Here, we consider the edge $e$ as spatial guidance, which is derived from Canny algorithm $\psi$:
\begin{equation}
    e =\psi(x_s, T)
\end{equation}
where $T$ is the Canny threshold and produced based on annotation. Following the ControlNet \cite{zhangAdding2023} strategy, we connect the locked DDPM and its trainable copy by zero convolution layers, and decode the edge into the model to learn spatial guidance. The learning objective of the entire edge-guided diffusion model $f_\theta$ is defined as:
\begin{equation}
L_{dif}=\mathbb{E}_{x_s,t,e,\epsilon\sim\mathcal{N}(0,1)}\Big[\left\|\epsilon-\epsilon_f\left(x_s^t,t,e\right)\right\|\Big]
\end{equation}
where $\epsilon$ is real noise and $\epsilon_f$ is estimated noise. $x_s^t$ is the noisy image, obtained by progressively adding the random noise $t$ times to a source image $x_s$. The optimization of the edge-guided diffusion model involves minimizing the discrepancy between real and estimated noises.

% The conditional diffusion model aligns the target image to match the style of the source domain while preserving the original spatial structure. Our diffusion model is fundamentally a conditional denoising diffusion probabilistic model \cite{hoDenoising2020} (DDPM) and used the design of ControlNet \cite{zhangAdding2023} to control the generation of the DDPM. 
% It consisted of two main components: a locked U-Net and a trainable copy of its decoder and middle block. These two components were connected in the decoder using  "zero convolution" layers —  $1\times1$ convolution layers with weights and biases initialized to zeros, and these layers were disconnected to improve the training speed. We employed the U-Net as the base network, which comprised of 6 blocks in the encoder, a middle block, and 6 blocks in the skip-connected decoder. Each block consisted of ResNet and downsample layers, and the multi-head attention layer was added to low-resolution blocks.

% Evidential Uncertainty estimation for segmentation
% 或者 xx segmentation with uncertainty estimation 
\subsection{Uncertainty segmentation model}
The uncertainty segmentation model is employed to produce segmentation masks, with the uncertainty estimation being utilized for pseudo-label refinement. We use the basic U-Net \cite{ronnebergerUNet2015} for the segmentation task and the learning objective consist of the commonly used cross-entropy loss and Dice loss: 
\begin{equation}
\label{seg_l}
    L_{seg} = \sum_{n=1}^C-y_s^n\log(\hat{y}) + 1-\frac{2*y_s*\hat{y}+1}{y_s+\hat{y}+1}
\end{equation}
where $\hat{y}=f_\theta(x_s)$ is the predicted mask. 

In our work, we expanded the U-Net architecture by introducing an additional branch aimed at estimating prediction uncertainty. Importantly, this extension seamlessly integrates into the existing U-Net structure without necessitating any fundamental alterations. This new branch is dedicated to directly assessing uncertainty by predicting the parameters of the Normal-Inverse-Gamma (NIG) distribution \cite{aminiDeep2020}. The NIG distribution serves as a conjugate prior distribution for the source domain distribution. We consider the relationship as follows: $y_s$ follows a normal distribution $\mathcal{N}(\mu,\sigma^2)$, where $\mu$ follows $\mathcal{N}(\gamma,\sigma^2\omega^{-1})$ and $\sigma^2$ follows $\Gamma^{-1}(\alpha, \beta)$, where $\Gamma^{-1}$ represents the gamma function. This formulation characterizes the distribution of $y_s$ as an NIG distribution with parameters $\gamma$, $\omega$, $\alpha$, and $\beta$.

\begin{equation}
\label{NIG}
    p(\mu, \sigma^2|\gamma,\omega,\alpha,\beta)=\frac{\beta^{\alpha}\sqrt{\omega}}{\Gamma(\alpha)\sqrt{2\pi\sigma^2}}\left(\frac{1}{\sigma^2}\right)^{\alpha+1}\exp\left\{-\frac{2\beta+\omega(\gamma-\mu)^2}{2\sigma^2}\right\}
\end{equation}
where $\gamma\in \mathbb{R}$, $\omega>1$ and $\beta>0$. In the training phase, we apply the following negative log-likelihood loss to account for the NIG distribution
\begin{equation}
\label{NIG_l}
    L_{NIG}=\frac{1}{2}\log\left(\frac{\pi}{\omega}\right)-\alpha\log(\Omega)+(\alpha+\frac{1}{2})\log((y-\gamma)^2\omega+\Omega)+\log(\Theta)
\end{equation}
where $\Omega=2\beta(1+\omega)$ and $\Theta=\left(\frac{\Gamma(\alpha)}{\Gamma(\alpha+\frac{1}{2})}\right)$. Furthermore, we introduce a regularization term $L_R=|y_i-\gamma|\cdot(2\omega+\alpha)$ into the total loss to penalize incorrect evidence
\begin{equation}
\label{un_l}
    L_{un}=L_{NIG}+\lambda L_{R}
\end{equation}
We introduce a coefficient $\lambda$ to balance the contributions of the two loss terms. Ultimately, we replace the deterministic output of the model with a NIG distribution $f_\theta(x_s)=NIG(\gamma, \omega, \alpha, \beta)$. The prediction $\hat{y}$ and the model uncertainty map $u$ are defined as:
\begin{equation}
    \hat{y}=\mathbb{E}(u)=\gamma
\end{equation}
\begin{equation}
    u =\frac{\beta}{\omega(\alpha-1)}
\end{equation}

\subsection{Reliable Source Approximation}
Due to domain shift, it is impossible to directly find an edge of the target image and obtain a high-quality source approximation through the edge-guided diffusion model learned from the source domain. Therefore, we generate multiple edges with $n$ thresholds $\{e_i|e_i=r(x_t,T_i),i=1,2,\dots,n\}$ for a target image $x_t$, and $n\times3$ images are generated under the guidance of edges. 
The key challenge is to find the optimal source approximation: the source-like and structure-preserved generated image with its reliable pseudo-label.

\subsubsection{Refine pseudo-label using uncertainty}
Domain shift can compromise the accuracy of pseudo-labels, potentially hindering overall model performance if the model is directly fine-tuned using them. To mitigate this issue, we employ uncertainty estimation to discern pixels with dependable pseudo-labels, subsequently utilizing only refined pseudo-labels for model supervision. Since the uncertainty segmentation model always exhibits high uncertainty at the edges of tumors when predicting generated images, we include the non-tumor pixels with high uncertainty at the edges in our analysis:
\begin{equation}
    \hat{y_{ij}}=\mathbbm{1}\{u_{ij}>T_{un}\}\oplus \hat{y_{ij}}\cup \hat{y_{ij}}
\end{equation}
where $T_{un}$ is the threshold of uncertainty, and $\mathbbm{1}\{u_{ij}>T_{un}\}$ find the high uncertainty prediction. The operation $\oplus$ finds non-tumor pixels with high uncertainty, and $\cup$ incorporates them into the predicted mask.

% 这两个符号说明一下是啥意思 圈圈加号和并集符号
\subsubsection{Prediction consistency} Under the guidance of same edge $e_i$, the spatial structures of all generated images $\{x_{i1},x_{i2},x_{i3}\}$ should be consistent if $e_i$ provides a precise spatial context. Further, their predictions $\{\hat{y_{i1}},\hat{y_{i2}},\hat{y_{i3}}\}$ should be similar and the prediction consistency $R_i$ is defined as: 
\begin{equation}
    R_i=1-\frac{\bigcap_{j=1}^{3} \hat{y_{ij}}}{\bigcup_{j=3}^{3} \hat{y_{ij}}}
\end{equation}
the edges with prediction consistency above $T_R$ are filtered out and best edge index is found by $argmin\{R[i] \ | \ R[i] \leq T_R, \ i \in [0, n]\}$. The final approximation $\{x_{ij},\hat{y_{ij}}\}$ is found based the smallest variance prediction in $\{\hat{y_{i1}},\hat{y_{i2}},\hat{y_{i3}}\}$.

\subsubsection{Batch-based and centralized fine-tuning} 
% We provide two . Finally, all $x_t$, $y_{ij}$, and $\hat{y_{ij}}$ are used to fine-tune the uncertainty segmentation model for adaptation.
We offer two fine-tuning methods here. Batch-based fine-tuning solely utilizes the current batch to train the model and directly predicts the results for the current batch. Batch-based fine-tuning only trains one epoch, and all batches are processed once. Centralized fine-tuning requires aggregating all batches and conducts multiple epochs of training.

\section{Experiment}
\subsection{Experimental Setup}

\subsubsection{Dataset and evaluation metrics}
We validate our SFUDA approach on the public vestibular schwannoma dataset including ceT1 and hrT2 MRI sequences from 242 patients \cite{shapeyArtificial2019}. We consider ceT1 MRI as the source domain and hrT2 as the target domain.
Following the data split in \cite{shapeyArtificial2019} and slicing the MRI volume, we finally obtain 1599 source training images, 1493 target training images, and 423 target testing images. In the evaluation process, the Dice similarity coefficient (Dice) and the average symmetric surface distance (ASSD) are employed to quantitatively assess the segmentation outcomes. \cite{shapeyArtificial2019}.

% \usepackage{multirow}

% \begin{table}[!t]
% \centering
% \caption{Quantitative results of comparison with different methods. The best scores are highlighted.}
% \begin{tabular}{c|c|c|cc}
% \hline\hline
% Strategy & \ \ \ \  Method\ \ \ \  & \ \ \ \ Venue \ \ \ \ & Dice (\%) $\uparrow$ & ASSD (mm) $\downarrow$\\ \hline\hline
% No adaptation & Source-only & - & 6.63 ± 1.85 & 2.10 ± 0.29 \\ \hline
% Supervised & Target-only& - & 84.49 ± 0.96 & 0.87 ± 0.06 \\ \hline
% \multirow{3}{*}{\begin{tabular}[c]{@{}c@{}}Entropy\\minimization\end{tabular}} & TENT\cite{wangTent2021} & ICLR'20 & 26.55 ± 5.69 & 9.54 ± 0.80 \\ \cline{2-5} 
%  & EATA\cite{niuEfficient2022} & PMLR'22 & 26.54 ± 5.70 & 9.53 ± 0.78 \\ \cline{2-5} 
%  & SAR\cite{niuStable2023} & ICLR'23 & 26.30 ± 5.80 & 9.51 ± 0.81 \\ \hline
% Pseudo-labeling & COTTA\cite{wangContinual2022} & CVPR'22 & 26.54 ± 5.80 & 9.42 ± 0.77 \\ \hline
% \multirow{3}{*}{\begin{tabular}[c]{@{}c@{}}Source\\approximation\end{tabular}} & FSM\cite{yangSource2022} & MIA'22 & 56.12 ± 1.32 & 4.56 ± 0.54 \\ \cline{2-5}
%  & Ours (Batch-based) & - & 45.55 ± 3.01 & 8.23 ± 1.11 \\ \cline{2-5}
% & Ours (Centralized) & - & \textbf{77.83 ± 0.98} & \textbf{1.24 ± 0.11} \\ 
%  \hline\hline
% \end{tabular}
% \label{tab:comparison}
% \end{table}

\subsubsection{Implementation details}
The experiments are conducted on the PyTorch platform with an NVIDIA A100 GPU. We first train DDPM for 400 epochs using source images, and then update the entire edge-guided diffusion model for 100 epochs simultaneously using source images and their edges. The diffusion model training uses the AdamW optimizer with a learning rate of 1e-4 and the batch size is 16. The sampler is Denoising Diffusion Implicit Models (DDIM) \cite{songDenoising2022} with 50 steps. We train the uncertainty segmentation model for 100 epochs and use the Adam optimizer with a learning rate of 1e-4 and a batch size of 32. Both the generated images and inputs of the uncertainty segmentation model are $320 \times 320$ resolutions. In the adaptation phase, we equally divide the range from 30 to 80 to obtain $n$ Canny thresholds. The hyper-parameters are set as $n=2$, $T_{un}=0.2$, $T_R=0.3$ through ablation study.

\begin{table}[t]
\centering
\caption{Quantitative results of comparison with different methods. The best scores of batch-based and centralized fine-tuning are highlighted.}
\begin{tabular}{c|c|c|cc}
\hline\hline
Strategy & Method & \ \ Fine-tuning\ \ & \ \ Dice (\%) $\uparrow$ \ \  &\ \   ASSD (mm)$\downarrow$ \ \ \\\hline\hline
No adaptation & Source-only & - & 6.63 ± 1.85 & 2.10 ± 0.29 \\ \hline
Supervised & Target-only& - & 84.49 ± 0.96 & 0.87 ± 0.06 \\ \hline
\multirow{6}{*}{\begin{tabular}[c]{@{}c@{}}Entropy\\      minimization\end{tabular}} & \multirow{2}{*}{TENT\cite{wangTent2021}} & Batch-based & 26.53 ± 4.73 & 9.56 ± 0.70 \\ \cline{3-5} 
 &  & Centralized & 26.55 ± 5.69 & 9.54 ± 0.80 \\ \cline{2-5} 
 & \multirow{2}{*}{EATA\cite{niuEfficient2022}} & Batch-based & 26.53 ± 4.73 & 9.55 ± 0.71 \\ \cline{3-5} 
 &  & Centralized & 26.54 ± 5.70 & 9.53 ± 0.78 \\ \cline{2-5} 
 & \multirow{2}{*}{SAR\cite{niuStable2023}} & Batch-based & 26.31 ± 5.82 & 9.52 ± 0.80 \\ \cline{3-5} 
 &  & Centralized & 26.30 ± 5.80 & 9.51 ± 0.81 \\ \hline
\multirow{2}{*}{\begin{tabular}[c]{@{}c@{}}Pseudo\\-labeling\end{tabular}} & \multirow{2}{*}{COTTA\cite{wangContinual2022}} & Batch-based & 26.52 ± 4.83 & 9.50 ± 0.60 \\ \cline{3-5}
 &  & Centralized & 26.54 ± 5.80 & 9.42 ± 0.77 \\ \hline
\multirow{3}{*}{\begin{tabular}[c]{@{}c@{}}Source\\approximation\end{tabular}} & FSM\cite{yangSource2022} & Centralized  & 56.12 ± 1.32 & 4.56 ± 0.54 \\ \cline{2-5}
 & \multirow{2}{*}{Ours}  & Batch-based & \textbf{45.55 ± 3.01} & \textbf{8.23 ± 1.11} \\ \cline{3-5}
 & & Centralized & \textbf{77.83 ± 0.98} & \textbf{1.24 ± 0.11}\\
 \hline\hline
\end{tabular}
\label{tab:comparison}
\end{table}

\begin{figure}[!t]
    \centering
    \includegraphics[width=\textwidth]{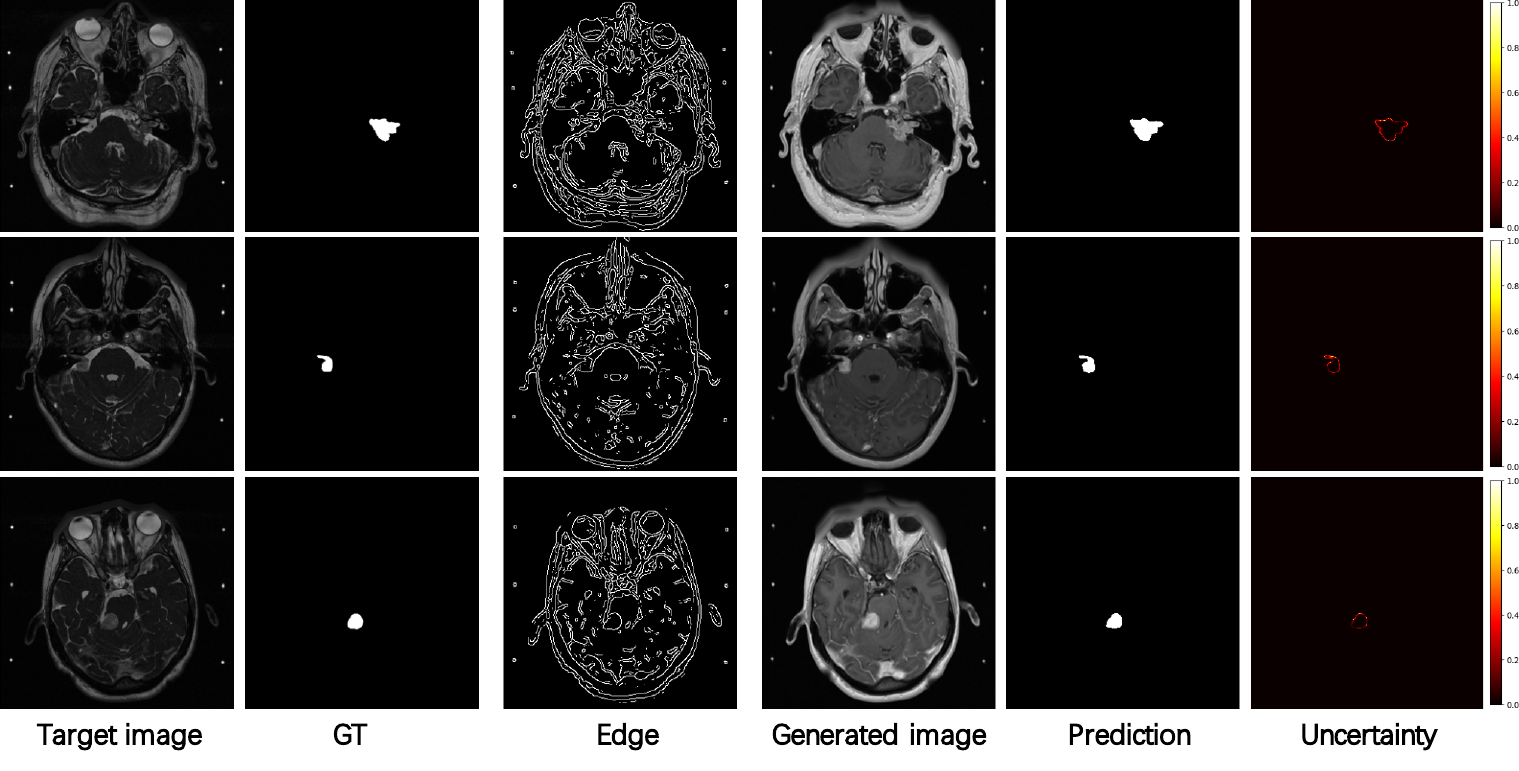}
    \caption{Visulazition of reliable source approximation and segmentation results. }
    \label{fig:vis}
\end{figure}

\subsection{Experimental Results}
\subsubsection{Quantitative comparison}
In our experiments, the "source-only" method refers to training a model solely on the source domain and evaluating it directly on the target domain without any adaptation. Conversely, the "target-only" method involves training and testing exclusively on the target domain. We conducted comparisons between our approaches and recent SFUDA methods employing various strategies, with many of these methods employing dual fine-tuning techniques.

Table \ref{tab:comparison} presents quantitative comparison results. There is a huge performance gap between source-only and target-only methods due to the domain shift between ceT1 and hrT2 MRI. The Dice score and ASSD of our method significantly outperform the entropy minimization and pseudo-labeling methods in two fine-tuning methods (Dice improvements $>45\%$ and $>15\%$, ASSD improvements $> 8 mm$ and $>1 mm$). These improvements can be attributed to the source approximation ability in tackling large domain shifts. Our method still outperforms the other source approximation FSM by a margin of 21.71\% in Dice and 3.32 mm in ASSD. All entropy minimization methods achieve similar results, which means that the latest improvements over TENT have not been very effective for VS segmentation tasks. Moreover, our method demonstrates greater stability compared to other approaches, since it achieves the least performance variance. Figure \ref{fig:vis} shows the best source approximations and their segmentation results, the edge-guided approximations are source-like and structure-preserved.

\subsubsection{Hyper-parameter Sensitivity Analysis}
We further investigate the impact of different hyper-parameters through source approximation and model performance. The quality of source approximation could be measured by the Dice score between pseudo-labels and ground truth. 

Table \ref{tab:Tr} presents the results with different prediction consistency thresholds $T_R$. It presents that both too-loose and too-tight thresholds would cause a drop in performance: a loose threshold of 0.9 generates a large amount of low-quality data, which is detrimental to model fine-tuning; a tight threshold of 0.1 yields better-quality data but in smaller quantities, potentially preventing the model from acquiring new knowledge. 
Table \ref{tab:Tun} presents the results with different uncertainty threshold $T_{un}$. We find that a more relaxed $T_{un}$ is more beneficial to results, as too stringent $T_{un}$ may lead to an excessive number of pixels being added to the predicted mask and damages the pseudo-label quality. We also find that $T_{un}$ only affects the approximation quality but does not impact the quantity.
Table \ref{tab:N} presents the results with different numbers of edges $n$. Excessive edges could lead to a surge in low-quality approximations and noisy pseudo-labels, resulting in decreased performance.

\begin{table}[h]
\centering
\caption{Ablation study for hyperparameter $T_R$.}
\begin{tabular}{c|c|c|c|c}
\hline\hline
\multirow{2}{*}{$\quad T_R\quad $} & \multicolumn{2}{c|}{Approximation} & \multicolumn{2}{c}{Performance} \\ \cline{2-5} 
& \multicolumn{1}{c|}{Quality (\%)} & Quantity (n) & \multicolumn{1}{c|}{Dice (\%)} & \multicolumn{1}{c}{ASSD (mm)} \\ \hline\hline
 % \hline
0.1 & \textbf{86.86} & 278 & 77.47 & 1.49 \\
0.3 & 78.58 & 706 & \textbf{77.63} & 1.53 \\
0.5 & 73.85 & 910 & 76.55 & \textbf{1.46} \\
0.7 & 69.88 & 1064 & 73.11 & 1.62 \\
0.9 & 65.33 & \textbf{1226} & 72.06 & 1.61 \\
\hline\hline
\end{tabular}
\label{tab:Tr}
\end{table}

\begin{table}[h]
\centering
\caption{Ablation study for hyperparameter $T_{un}$.}
\begin{tabular}{c|c|c|c|c}
\hline\hline
\multirow{2}{*}{$\quad T_{un}\quad $} & \multicolumn{2}{c|}{Approximation} & \multicolumn{2}{c}{Performance} \\ \cline{2-5} 
& \multicolumn{1}{c|}{Quality (\%)} & Quantity (n) & \multicolumn{1}{c|}{Dice (\%)} & \multicolumn{1}{c}{ASSD (mm)} \\ \hline\hline
0.002 & 72.66 & 910 & 74.95 & 1.47\\
0.02 & 73.85 & 910 & 75.59 & \textbf{1.33} \\
0.2 & \textbf{73.92} & 910 & \textbf{76.32} & 1.41 \\
\hline\hline
\end{tabular}
\label{tab:Tun}
\end{table}

\begin{table}[h]
\centering
\caption{Ablation study for hyperparameter $n$.}
\begin{tabular}{c|c|c|c|c}
\hline\hline
\multirow{2}{*}{$\quad n\quad $} & \multicolumn{2}{c|}{Approximation} & \multicolumn{2}{c}{Performance} \\ \cline{2-5} 
& \multicolumn{1}{c|}{Quality (\%)} & Quantity (n) & \multicolumn{1}{c|}{Dice (\%)} & \multicolumn{1}{c}{ASSD (mm)} \\ \hline\hline
2 & \textbf{76.77} & 786 & \textbf{78.53} & \textbf{1.17} \\
4 & 73.85 & 910 & 75.59 & 1.33 \\
8 & 68.72 & \textbf{1107} & 73.21 & 1.35\\
\hline\hline
\end{tabular}
\label{tab:N}
\end{table}

\section{Conclusion}
This study introduces a novel SFUDA method tailored for the vestibular schwannoma MRI segmentation. We implement a novel source-free source approximation method via edge-guided image translation. The uncertainty segmentation model and prediction consistency are introduced to obtain Reliable Source Approximation results. Experimental results on cross-sequence MRI image segmentation demonstrate that our method outperfroms state-of-the-art SFUDA approaches.

\bibliographystyle{IEEEtran}
\bibliography{ref}
\end{document}